\documentclass[]{ws-ijmpb}
\usepackage[]{graphicx}                                                                  

\newcommand{\ud}{\mathrm{d}}
  
\begin{document}

\markboth{Fedorova, Zeitlin}                                                                
{Pattern Formation in Quantum Ensembles}                                                                              
\catchline{}{}{}{}{}   
                                                         
\title{PATTERN FORMATION IN QUANTUM ENSEMBLES}

\author{ANTONINA N. FEDOROVA and MICHAEL G. ZEITLIN}                                        
\address{IPME RAS, St.~Petersburg,\\                                                        
 V.O. Bolshoj pr., 61, 199178, Russia\\                                                     
http://www.ipme.ru/zeitlin.html,                                                          
http://www.ipme.nw.ru/zeitlin.html,\\                                                       
E-mail: zeitlin@math.ipme.ru, anton@math.ipme.ru}                                           
                                                                                            
\maketitle                                                                                  
\begin{history}                                                                             
%\received{Day Month Year}                                                                   
%\revised{Day Month Year}                                                                    
%\accepted{(Day Month Year)}                                                                
\comby{(xxxx) }                                                                       
\end{history}

%%%%%%%%%%%%%%%%%%%%%%%%%%%%%%%%%%%%%%%%%%%%%%%%%%%%%%%%%%%%%%%%%%%%%%%%%%%%%%%%%

\begin{abstract}  

We present a family of methods, analytical and
numerical, which can describe behaviour in 
(non) equilibrium ensembles, both classical and quantum, especially in
the complex systems, where the standard approaches cannot
be applied. 
We demonstrate the creation of nontrivial (meta) stable
states (patterns), localized, chaotic, entangled or decoherent, 
from basic localized modes
in various collective models arising from the 
quantum hierarchy 
of Wigner-von Neumann-Moyal-Lindblad equations, 
which are the result of ``wignerization'' procedure
of classical BBGKY hierarchy. We present the
explicit description of internal quantum dynamics by means of 
exact analytical/numerical computations. 

\end{abstract} 
  

\section{Localized Modes (``continuous qudits''): Why Need We Them?}

It is widely known that the currently available experimental techniques (and, 
apparently, those which will become avaiable in the nearest future) in the area of 
quantum physics as a whole and in that of quantum computations in particular, as well 
as the present level of understanding of phenomenological models, outstripped
the actual level of mathematical/theoretical description\cite{1}.
Considering, for example, the problem of describing the realizable states
(Refs.~\refcite{2}--\refcite{7}),
one 
should not expect that planar waves and (squeezed) gaussian coherent states 
would be enough to characterize such complex systems as qCPU (quantum Central Processor Unit)-like devices.
Complexity of the set of relevant states, including entangled (chaotic) 
states is still far from being clearly understood and moreover from being realizable.
As a starting point for our approach let us consider the following well-known 
example of  
GKP (Gottesman, Kitaev, Preskill)\cite{8} scheme with DV (Discrete Variables)/qubit 
(with finite-dimensional code space 
embedded in the infinite-dimensional Hilbert space) or CV (Continuous Variables) for (optical) 
quantum computations,
containing as a part (optical) nonlinearities, described by 
Kerr interaction or more general polynomial Hamiltonians which are needed to realize the   
state preparation and provide the process of CV quantum computation\cite{8}. 
It is an important example because: 

{\bf(a)} its classical counterpart is described by polynomial Hamiltonians;

{\bf(b)} the proper qudits or building states (DV or CV) 
are well localized (but not well-defined mathematically, as we 
shall explain later).

\noindent One of the questions which motivated our approach is whether it is 
possible to keep {\bf(a)} and at the same time improve {\bf(b)}.
Our other motivations arise from the following 
general questions:

{\bf(A)} How can we represent well localized and reasonable state in mathematically correct form?

{\bf(B)} Is it possible to create entangled and other relevant states by means of these new building blocks?

\noindent In GKP scheme unphysical and not clearly defined mathematically logical qubit states   
are represented via infinite series of $\delta$ functions: 
$
\vert 0>=\sum^\infty_{s=-\infty}\delta(x-2s\sqrt\pi)|x>, 
\vert 1>=\sum^\infty_{s=-\infty}\delta(x-(2s+1)\sqrt\pi)|x>
$
and approximated by the set of gaussian envelopes:
$
<x\vert 0>=N_0\sum^{+\infty}_{-\infty}e^{-1/2(2sk\sqrt\pi)^2}
e^{-1/2(\frac{x-2s\sqrt\pi)}{\triangle})^2}$,
$
<x\vert 1>=N_1\sum^{+\infty}_{-\infty}e^{-1/2(2s+1)k\sqrt\pi)^2}
e^{-1/2(\frac{x-(2s+1)\sqrt\pi)}{\triangle})^2}.
$

Due to numerous mathematical and computational
reasons, some of which are described below, 
such and related choices cannot be appropriate neither as a starting point on 
the route to the real qCPU device nor as a satisfactory theoretical 
description.
So, it would appear that a first step in this direction is
to find a reasonable extension of understanding of the quantum dynamics
as a whole.
One needs to sketch up the underlying ingredients of 
the theory (spaces of  states, observables, measures,
classes of smoothness, quantization set-up etc) in an attempt 
to provide the maximally extendable
but at the same time really calculable and realizable 
description of the dynamics of quantum world.
The general idea is rather simple: it is well known that the
idea of ``symmetry'' is the key ingredient of any
reasonable physical theory from classical (in)finite dimensional
(integrable) Hamiltonian
dynamics to different sub-planckian models based on
strings (branes, orbifolds etc.)
During the last century kinematical, dynamical and hidden symmetries
played the key role in our understanding of physical process.
Roughly speaking, the representation theory of underlying 
symmetry (classical or quantum, groups or (bi)algebras, finite 
or infinite dimensional, continuous 
or discrete) is a proper instrument for description of proper (orbital) dynamics.
A starting point for us is a possible model for (continuous) ``qudit'' with subsequent 
description of the whole zoo of possible realizable (controllable) states/patterns 
which may be useful from the point of view of quantum 
experimentalists and engineers.
The proper representation theory is well known as ``local nonlinear harmonic analysis'',
in particular case of simple underlying symmetry--affine group--aka wavelet analysis.
From our point of view the advantages of such approach are as follows:

\noindent{\bf i)} natural realization of localized states in any proper functional realization of 
(Hilbert) space of states,

\noindent{\bf ii)} hidden symmetry of chosen realization of proper functional model provides 
the (whole) spectrum of possible states via the so-called 
multiresolution decomposition.

So, indeed, the hidden symmetry (non-abelian affine group in the simplest case) 
of the space of states via proper representation theory generates the physical spectrum
and this procedure depends on the choice of the functional realization of the space of states.  
It explicitly demonstrates that the structure and properties of the functional realization of the
space of states are the natural properties of physical world at the same level of importance as
a particular choice of Hamiltonian, or the equation of motion, or the action principle (variational method).      
At the next step we need to consider the consequences of 
our choice i), ii) for the algebra of observables.
In this direction one needs to mention the class of operators we are interested in to
present proper description for a class of maximally 
generalized but reasonable class of problems.
It seems that these must be pseudodifferential operators, especially 
if we underline that in the spirit of points i), ii) above we need to take Wigner-Weyl framework 
for constructing basic quantum equations of motions.
It is obvious, that consideration of symbols of operators instead of operators themselves
is the starting point as for the mathematical theory of pseudodifferential operators as for 
quantum dynamics formulated in the language of Wigner-like equations.
It should be noted that in such picture 
we can naturally include the effects 
of self-interaction (``quantum non-linearity'') on the way of 
construction and subsequent analysis of nonlinear quantum models.
So, our consideration will be in the framework of (Nonlinear) Pseudodifferential Dynamics
($\Psi DOD$).
As a result of i), ii), we'll have:

\noindent{\bf iii)} most sparse, almost diagonal, representation for a wide 
class of operators included in the set-up of the whole problems.

It's possible by using the 
so-called Fast Wavelet Transform representation for algebra of observables.

Then points i)--iii) provide us by

\noindent{\bf iv)} natural (non-perturbative) multiscale decomposition 
for all dynamical quantities, as states as observables. 

The simplest case we will have, obviously, in Wigner-Weyl representation.
Existence of such internal multiscales 
with different dynamics at each scale and transitions, interactions,
and intermittency between scales demonstrates that quantum mechanics, despite  its
linear structure, is really a serious part of physics from the mathematical point of view. 
It seems, that well-known underlying quantum complexity is a result of transition 
by means of (still rather unclear) procedure of quantization
from complexity related to nonlinearity of classical counterpart to the rich 
pseudodifferential (more exactly, microlocal) structure on the quantum side.

We divide all possible configurations
related to possible solutions of our quantum equation 
of motion (Wigner-like equations, mostly)
into two classes: 

{\bf (a)} standard solutions;
{\bf (b)} controllable solutions (solutions with prescribed qualitative type of behaviour).

Anyway, the whole zoo of solutions consists of possible patterns, including
very important ones from the point of view of underlying physics:

\noindent{\bf v)} localized modes (basis modes, eigenmodes) and 
constructed from them chaotic or entangled,
decoherent (if we change Wigner equation for (master) Lindblad one) patterns.

It should be noted that these bases modes are nonlinear in contrast with usual ones
because they come from (non) abelian generic group while the usual Fourier (commutative) 
analysis starts from $U(1)$ abelian modes (plane waves).
They are really ``eigenmodes'' but in sense of decomposition of representation of the underlying
hidden symmetry group which generates the multiresolution decomposition.
The set of patterns is built from these modes by means of variational procedures 
more or less standard in mathematical physics.
It allows to control the convergence from one side but, what is more important,

\noindent{\bf vi)} to consider the problem of the control of patterns (types of behaviour) on the level
of reduced (variational) algebraical equations.

We need to mention that it is possible to change the 
simplest generic group of hidden internal symmetry 
from the affine (translations and dilations) to much more general, 
but, in any case, this generic 
symmetry will produce the proper 
natural high localized eigenmodes, as well as the decomposition 
of the functional realization
of space of states into the proper orbits; 
and all that allows to compute dynamical consequence 
of this procedure, i.e. pattern formation, and, as a result, 
to classify the whole spectrum of proper states.

For practical reasons controllable patterns (with prescribed behaviour) are the most useful.
We mention the so-called waveleton-like pattern which we regard as the most important one.
We use the following allusion in the space of words:

\{waveleton\}:=\{soliton\}  $\bigsqcup$   \{wavelet\}

It means:

\noindent{\bf vii)} waveleton $\approx$ (meta)stable localized (controllable) pattern

To summarize, the approach described below allows one 

\noindent{\bf viii)} to solve wide classes of general 
$\Psi DOD$ problems, including generic for quantum physics Wigner-like equations, and

\noindent{\bf ix)} to present the analytical/numerical realization for physically interesting patterns.

We would like to emphasize the effectiveness of numerical realization of this program
(minimal complexity of calculations) as additional advantage.
So, items i)-ix) point out all main features of our approach, Refs.~\refcite{2}--\refcite{7}.

\section{Motivations}
\subsection{Class of Models}

Here we describe a class of problems which can be analysed by methods described in Introduction.
We start from individual dynamics and finish by (non)-equilibrium ensembles.
All models belong to the $\Psi DOD$ class and can be described by finite or infinite 
(named hierarchies in such cases) system of 
$\Psi DOD$ equations:

\begin{itemize}

\item[a).]  Individual classical/quantum 
mechanics ($cM/qM$): linear/nonlinear; $\{cM\}\subset\{qM\}$, 
$\ast$ - quantized for the class of polynomial Hamiltonians
$
H(p,q,t)=\sum_{i.j}a_{ij}(t)p^iq^j.
$
\item[b).] QFT-like models in framework of the second quantization (dynamics in Fock spaces).
\item[c.)] Classical (non) equilibrium ensembles via BBGKY Hierarchy 
(with reductions to different forms of Vlasov-Maxwell/Poisson equations). 
\item[d.)] Wignerization of a): Wigner-Moyal-Weyl-von Neumann-Lindblad.
\item[e.)] Wignerization of c): Quantum (Non) Equilibrium Ensembles.

\end{itemize}

Important remarks: points a)-e) are considered in $\Psi DO$ 
picture of  (Non)Linear $\Psi DO$ Dynamics (surely, all $qM\subset\Psi DOD$);
dynamical variables/observables are the symbols of operators or functions; in case
of ensembles, the main set of dynamical variables consists 
of partitions (n-particle partition functions).

\subsection{Effects we are interested in}

\begin{romanlist}
\item Hierarchy of internal/hidden scales (time, space, phase space).
\item Non-perturbative multiscales: 
from slow to fast contributions,
from the coarser to the finer level of resolution/de\-composition.
\item Coexistence of hierarchy of multiscale dynamics with transitions between scales.
\item Realization of the key features of the complex quantum 
world such as the existence of chaotic and/or entangled 
states with possible destruction in ``open/dissipative'' regimes due to interactions with
quantum/classical environment and transition to decoherent states.

\end{romanlist}

At this level we may interpret the effect of mysterious entanglement or ``quantum interaction''  
as a result of simple interscale interaction or intermittency (with allusion to hydrodynamics),
i.e. the mixing of orbits generated by multiresolution representation of hidden underlying symmetry.
Surely, the concrete realization of such a symmetry is a natural physical property 
of the physical model as well as the space of representation and its proper functional 
realization.
So, instantaneous interactions (or transmission of ``quantum bits'' or ``teleportation'') 
materialize not in the physical space-time variety but in the space of representation of hidden symmetry
along the orbits/scales constructed by proper representations. 
Dynamical/kinematical principles of usual space-time varieties, definitely, do not cover
kinematics of internal quantum space of state or, in more weak formulation, 
we still have not such explicit relations.

One additional important comment:
as usual in modern physics, we have the hierarchy of underlying symmetries; so
our internal symmetry of functional realization of space 
of states is really not more than kinematical, because  much more rich
algebraic structure, related to operator Cuntz algebra and quantum groups, is hidden inside.
The proper representations can generate much more interesting effects than ones described above.
We will consider it elsewhere but mention here only how it can be realized by the existing  
functorial maps between proper categories:

\{QMF\} $\longrightarrow$ Loop groups $\longrightarrow$ Cuntz operator algebra 
$\longrightarrow$ Quantum Group structure,
where \{QMF\} are the so-called quadratic mirror filters generating the realization of 
multiresolution decomposition/representation in any functional space; loop group
is well known in many areas of physics, e.g. soliton theory, strings etc, 
roughly speaking, its algebra coincides with Virasoro algebra; Cuntz operator algebra
is universal $C^*$ algebra generated by N elements with two relations between them;
Quantum group structure (bialgebra, Hopf algebra, etc) is well known in many areas 
because of its universality.
It should be noted the appearance of natural Fock structure inside this functorial sequence
above with the creation operator realized as some generalization of Cuntz-Toeplitz isometries.
Surely, all that can open a new vision of old problems and bring new possibilities. 

We finish this part by the following qualitative definitions of key objects (patterns). 
Their description and understanding 
in different physical models is our main goal in this direction. 

\begin{itemlist}
\item
By localized states (localized modes) 
we mean the building blocks for solutions or generating modes which 
are localized in maximally small region of the phase (as in c- as in q-case) space.

\item
By an entangled/chaotic pattern we mean some solution (or asymptotics of solution) 
which has random-like distributed energy (or information) spectrum in a full domain of definition. 
In quantum case we need to consider additional entangled-like patterns, roughly speaking,
which cannot be separated into pieces of sub-systems.

\item
By a localized pattern (waveleton) 
we mean (asymptotically) (meta) stable solution localized in a 
relatively small region of the whole phase space (or a domain of definition). 
In this case the energy is distributed during some time (sufficiently large) 
between only a few  localized modes (from point 1). 
We believe it to be a good model for plasma in a fusion state (energy confinement)
or a model for quantum continuous ``qubit'' or a result of the decoherence process
in open quantum system when the full entangled state degenerates 
into localized (quasiclassical) pattern.

\end{itemlist}

\subsection{Methods}

\begin{romanlist}

\item Representation theory of internal/hidden/underlying symmetry,
Kinematical, Dynamical, Hidden.

\item Arena (space of representation): proper functional realization of (Hilbert) space of states.

\item
Harmonic analysis on (non)abelian group of internal symmetry.
Local/Nonlinear (non-abelian) Harmonic Analysis 
(e.g, wavelet/gabor etc. analysis) instead of linear non-localized $U(1)$ Fourier analysis.
Multiresolution (multiscale) representation.
Dynamics on proper orbit/scale (inside the whole hierarchy of multiscales) in functional space.
The key ingredients are 
the appearance of multiscales (orbits) and the existence of high-localized 
natural (eigen)modes \cite{9}.

\item
Variational formulation (control of convergence, reductions 
to algebraic systems, control of type of behaviour).

\end{romanlist}

\section {Set-up/Formulation} 

Let us consider the following generic $\Psi$DOD dynamical problem

$$
L^j\{Op^i\}\Psi=0,
$$
described by a finite or infinite number of 
equations which include general classes of operators
$Op^i$ such as differential, integral, pseudodifferential etc

Surely, all Wigner-like equations/hierarchies are inside.

The main objects are:

\begin{romanlist}
\item (Hilbert) space of states, $H=\{\Psi\}$, with a proper functional 
realization, e.g.,: $L^2$, Sobolev, Schwartz,
$C^0$, $C^k$, ... $C^\infty$, ...; 
Definitely, $L^2(R^2)$, $L^2(S^2)$, $L^2(S^1\times S^1)$, $L^2(S^1\times S^1\ltimes Z_n)$
are different objects proper for different physics inside.
     
\item
Class of smoothness. The proper choice determines natural consideration of dynamics 
with/without Chaos/Fractality property.

\item Decompositions 
$$
\Psi\approx\sum_ia_ie^i
$$ 

via high-localized bases (wavelet families, generic wavelet packets etc), 
frames, atomic decomposition
(Fig. ~1) with the following 
main properties:
(exp) control of convergence, maximal rate of convergence  for any $\Psi$ in any $H$.

\begin{figure}
\begin{center}
\begin{tabular}{c}
\includegraphics*[width=60mm]{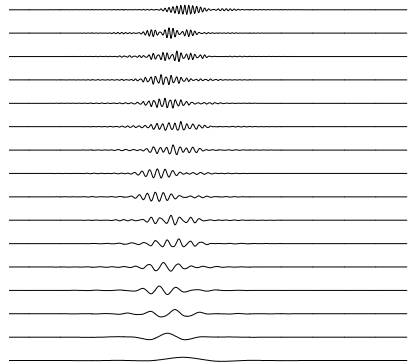}
\end{tabular}
\end{center}
\caption{Localized modes.}
\end{figure}

\item Observables/Operators (ODO, PDO, $\Psi$DO, SIO,...,
Microlocal analysis of Kashiwara-Shapira (with change from functions to sheafs))
satisfy the main property -- the matrix representation in localized bases
$$
<\Psi|Op^i|\Psi>
$$

is maximum sparse: 

\begin{displaymath}
\left(\begin{array}{cccc}
D_{11} & 0 &0 & \ldots\\
0    & D_{22} & 0 & \ldots\\
0    & 0    & D_{33} & \ldots\\
\vdots & \vdots & \vdots & \ddots
\end{array} \right).
\end{displaymath}

This almost diagonal structure is provided by the so-called Fast Wavelet Transform.

\item Measures: multifractal wavelet measures $\{\mu_i\}$ together with the class of smoothness
are very important for analysis of complicated analytical behaviour.

\item Variational/Projection methods, from Galerkin to 
Rabinowitz minimax, Floer (in symplectic case of Arnold-Weinstein curves with
preservation of Poisson/symplectic structures).
Main advantages are the reduction to algebraic systems, which provides a tool for the smart   
subsequent control of behaviour and control of convergence.

\item Multiresolution or multiscale decomposition, $MRA$ (or wavelet microscope) 
consists of the understanding and choosing of
   \begin{itemize}
   \item[1).] (internal) symmetry structure, e.g., 
affine group = \{translations, dilations\} or 
many others; construction of
   \item[2).] representation/action of this symmetry on $H=\{\Psi\}$.
    
As a result of such hidden coherence together with using point vi) we'll have:
     \begin{itemize}
     \item[a).] LOCALIZED BASES  $\qquad$ b). EXACT MULTISCALE DECOMPOSITION
with the best convergence properties and real evaluation of the rate of convergence
via proper ``multi-norms''.
      \end{itemize}
      \end{itemize}
Figures~2, 3, 5, 6 demonstrate MRA decompositions for one- and multi-kicks while Figures
4 and 7 present the same for the case of the generic simple fractal model, Riemann--Weierstrass
function [9].

\item Effectiveness of proper numerics: CPU-time, HDD-space, minimal complexity of
algorithms, and (Shannon) entropy of calculations are provided by points i)-vii) above.

\item Quantization via $\ast$ star product or Deformation Quantization.
\end{romanlist}

The corresponding class of individual Hamiltonians has the form
\begin{eqnarray}
\hat{H}(\hat{p},\hat{q})=\frac{\hat{p}^2}{2m}+U(\hat{p},\hat{q}),
\end{eqnarray}
where $U(\hat{p}, \hat{q})$ is an arbitrary polynomial 
function on $\hat{p}$, $\hat{q}$, 
and plays the key role in many areas of physics \cite{1}. 
Our starting point is the general point of view of a deformation 
quantization approach at least on
the naive Moyal/Weyl/Wigner level.
The main point of such approach is based on ideas from \cite{1}, which allow  
to consider the algebras of quantum observables as the deformations
of commutative algebras of classical observables (functions). 
So, if we have  the classical counterpart of 
Hamiltonian (1) as a model for classical dynamics 
and the Poisson manifold $M$ (or symplectic 
manifold or Lie coalgebra, etc) as the corresponding phase space, 
then for quantum calculations we need first of all to find
an associative (but non-commutative) star product 
 $*$ on the space of formal power series in $\hbar$ with
coefficients in the space of smooth functions on $M$ such that
$
f * g =fg+\hbar\{f,g\}+\sum_{n\ge 2}\hbar^n B_n(f,g), 
$
where
$\{f,g\}$
is the Poisson brackets, $B_n$ are bidifferential operators.
In this paper we consider the calculations of the Wigner functions
$W(p,q,t)$ (WF) corresponding
to the classical polynomial Hamiltonian $H(p,q,t)$ as the solution
of the Wigner-von Neumann equation \cite{1}:
\begin{eqnarray}
i\hbar\frac{\partial}{\partial t}W = H * W - W * H
\end{eqnarray}
and related Wigner-like equations for different ensembles.
According to the Weyl transform, a quantum state (wave function or density 
operator $\rho$) corresponds
to the Wigner function, which is the analogue in some 
sense of classical phase-space distribution \cite{1}.
Wigner equation (2) is a result of the Weyl transform 
or ``wignerization'' of von Neumann equation for density matrix.

Finally, such Variational-Multiscale approach based on points i)-ix) 
provides us by the full ZOO of PATTERNS: 
LOCALIZED, CHAOTIC/ENTANGLED, etc. 

In next Sections we will consider 
details for important cases of Wigner-like equations.

We present the explicit analytical construction for solutions of
c- and q-hierarchies and their important reductions starting from quantization 
of c-BBGKY (Born-Bogolyubov-Green-Kirkwood-Yvon) hierarchy, 
which is based 
on tensor algebra extensions of multiresolution
representation for states and observables and variational formulation.
We give explicit representation for hierarchy of n-particle 
reduced distribution functions 
in the base of
high-localized generalized coherent (regarding underlying generic symmetry 
(affine group in the simplest case)) 
states given by polynomial tensor algebra of our basis functions 
(wavelet families, wavelet packets), which 
takes into account
contributions from all underlying hidden multiscales
from the coarsest scale of resolution to the finest one to
provide full information about (quantum) dynamical process.
The difference between classical and quantum case is concentrated
in the structure of the set of operators included in the set-up and, surely,
depends on the method of quantization.
But, in the naive Wigner-Weyl approach for quantum case the symbols of operators  
play the same role as usual functions in classical case. 
In some sense,
our approach for ensembles (hierarchies) resembles Bogolyubov's one and related approaches 
but we don't use any perturbation technique (like virial expansion)
or linearization procedures.
Most important, that 
numerical modeling in all cases shows the creation of
different internal (coherent)
structures from localized modes, which are related to stable (equilibrium) or 
unstable type of behaviour and corresponding pattern (waveletons) formation.

\section{BBGKY/Wigner Ensembles: from $c$- to $q$-cases}

We start from  set-up for kinetic BBGKY hierarchy 
as c-counterpart of proper q-hierarchy.
Let M be the phase space of ensemble of N particles ($ {\rm dim}M=6N$)
with coordinates
$
x_i=(q_i,p_i), \quad i=1,...,N,
q_i=(q^1_i,q^2_i,q^3_i)\in R^3,
p_i=(p^1_i,p^2_i,p^3_i)\in R^3,
q=(q_1,\dots,q_N)\in R^{3N}.
$
Individual and collective measures are: 
$
\mu_i=\ud x_i=\ud q_i\ud p_i,\quad \mu=\prod^N_{i=1}\mu_i
$ while
distribution function
$D_N(x_1,\dots,x_N;t)$
satisfies 
Liouville equation of motion for ensemble with Hamiltonian $H_N$ 
and normalization constraint.
Our constructions can be applied to the following general Hamiltonians:
\begin{eqnarray}
H_N=\sum^N_{i=1}\Big(\frac{p^2_i}{2m}+U_i(q)\Big)+
\sum_{1\leq i\leq j\leq N}U_{ij}(q_i,q_j),  
\end{eqnarray}
where potentials 
$U_i(q)=U_i(q_1,\dots,q_N)$ and $U_{ij}(q_i,q_j)$
are not more than rational functions on coordinates.
Let $L_s$ and $L_{ij}$ be the standard Liouvillean operators 
and
\begin{eqnarray}
F_s(x_1,\dots,x_s;t)=
V^s\int D_N(x_1,\dots,x_N;t)\prod_{s+1\leq i\leq N}\mu_i
\end{eqnarray}
be the hierarchy of reduced distribution functions, then,
after standard manipulations, we arrive to c-BBGKY hierarchy:
\begin{eqnarray}
\frac{\partial F_s}{\partial t}+L_sF_s=\frac{1}{\upsilon}\int\ud\mu_{s+1}
\sum^s_{i=1}L_{i,s+1}F_{s+1}.
\end{eqnarray}
So, the proper dynamical formulation is reduced to the
(infinite) set of equations for correlators/partition functions.
Then by using physical motivated reductions or/and during 
the corresponding cut-off procedure 
we obtain, 
instead of linear and pseudodifferential (in general case)
equations,
their finite-dimensional but nonlinear approximations with
the polynomial type of nonlinearities (more exactly, multilinearities).
To move from $c$- to $q$-case, let us start from the second quantized 
representation for an algebra of observables 
$A=(A_0,A_1,\dots,A_s,...)$
in the standard form
$
A=A_0+\int dx_1\Psi^+(x_1)A_1\Psi(x_1)+\dots+$$
(s!)^{-1}\int dx_1\dots$$ dx_s\Psi^+(x_1)\dots$$
\Psi^+(x_s)A_s\Psi(x_s)\dots$\\
$
\Psi(x_1)+\dots
$
N-particle Wigner functions
\begin{eqnarray}
W_s(x_1,\dots,x_s)&=&\int dk_1\dots dk_s{\rm exp}\big(-i\sum^s_{i=1}k_ip_i\big)
{\rm Tr}\rho\Psi^+\big(q_1-\frac{1}{2}\hbar k_1\big)\dots\\
& &\Psi^+\big(q_s-\frac{1}{2}\hbar k_s\big)\Psi\big(q_s+\frac{1}{2}\hbar 
k_s\big)\dots
\Psi\big(q_1+\frac{1}{2}\hbar k_s\big)\nonumber
\end{eqnarray}
allow us to consider them as some quasiprobabilities and provide useful bridge
between c- and q-cases:
\begin{equation}
<A>={\rm Tr}\rho A=\sum^{\infty}_{s=0}(s!)^{-1}\int\prod_{i=1}^s 
d\mu_iA_s(x_1,\dots,x_s)
W_s(x_1,\dots,x_s).
\end{equation}
The full description for quantum ensemble can be done by the whole hierarchy
of functions (symbols):
$
W=\{W_s(x_1,\dots,x_s), s=0,1,2\dots\}
$
So, we may consider the following q-hierarchy as the result of ``wignerization'' 
procedure for c-BBGKY one:
\begin{eqnarray}
\partial_tW_s(t,x_1,\dots,x_s)&=&\sum^s_{j=1}L_j^0W_s(x_1,\dots,x_s)+
\sum_{j<n}\sum^s_{n=1}L_{j,n}W_s(x_1,\dots,x_s)\\
&+&
\sum^s_{j=1}\int 
dx_{s+1}\delta(k_{s+1})L_{j,s+1}W_{s+1}(x_1,\dots,x_{s+1})\nonumber,
\end{eqnarray}
\begin{eqnarray}
&&L^0_j=-\big(\frac{i}{m}\big)k_jp_j,
\qquad
L_{j,n}=(i\hbar)^{-1}\int d\ell \tilde{V_l}\Bigg[{\rm 
exp}\Bigg(-\frac{1}{2}\hbar\ell
\big(\frac{\partial}{\partial p_j}-\frac{\partial}{\partial p_n}\big)\Bigg)-\nonumber\\
&&{\rm exp}\bigg(\frac{1}{2}\hbar\ell\big(\frac{\partial}{\partial p_j}-
\frac{\partial}{\partial p_n}\big)\Bigg)
\Bigg]{\rm exp}\Bigg(-\ell\big(\frac{\partial}{\partial 
k_j}-\frac{\partial}{\partial k_n}\big)\Bigg).
\end{eqnarray}
In quantum statistics the ensemble properties are described by the density
operator
\begin{equation}
\rho(t)=\sum_i w_i|\Psi_i(t)><\Psi_i(t)|, \quad \sum_iw_i=1.
\end{equation}
After Weyl transform we have the following
 de\-com\-position via partial Wigner functions 
$W_i(p,q,t)$ for the whole ensemble Wigner function:
\begin{equation}
W(p,q,t)=\sum_iw_iW_i(p,q,t), 
\end{equation}
where the partial Wigner functions

\begin{eqnarray}
W_n(q,p,t)\equiv\frac{1}{2\pi\hbar}\int\ud\xi{\rm exp}\Big(-\frac{i}{\hbar}p\xi\Big)
\Psi^*_n(q-\frac{1}{2}\xi,t)\Psi_n(q+\frac{1}{2}\xi,t)
\end{eqnarray}
are solutions of proper Wigner equations:

\begin{eqnarray}
\frac{\partial W_n}{\partial t}=-\frac{p}{m}\frac{\partial W_n}{\partial q}+
\sum^{\infty}_{\ell=0}\frac{(-1)^\ell(\hbar/2)^{2\ell}}{(2\ell+1)!}
\frac{\partial^{2\ell+1}U_n(q)}{\partial q^{2\ell+1}}
\frac{\partial^{2\ell+1}W_n}{\partial p^{2\ell+1}}.
\end{eqnarray}
Our approach, presented below, in some sense has allusion on the analysis of the following
standard simple model considered in Ref.~\refcite{1}.
Let us consider model of interaction of nonresonant atom with quantized electromagnetic field:
$
\hat{H}=\frac{\hat{p}_x^2}{2m}+U(\hat{x}),\qquad
U(\hat{x})=U_0(z,t)g(\hat{x})\hat{a}^+\hat{a},  
$
where potential $U$ 
depends on creation/annihilation operators and some polynomial on $\hat{x}$ 
operator function (or approximation)
$g(\hat{x})$.
It is possible to solve Schroedinger equation
$
i\hbar{\ud|\Psi>}/{\ud t}=\hat{H}|\Psi>
$
by the simple ansatz 
\begin{eqnarray}
|\Psi(t)>=\sum_{-\infty}^{\infty}w_n\int\ud x \Psi_n(x,t)|x>\otimes|n>,
\end{eqnarray}
which leads to the hierarchy of analogous equations with potentials created by 
n-particle Fock subspaces
\begin{eqnarray}
i\hbar\frac{\partial\Psi_n(x,t)}{\partial t}=\Big\{\frac{\hat{p}_x^2}{2m}+
 U_0(t)g(x)n\Big\}\Psi_n(x,t),
\end{eqnarray}
where
$\Psi_n(x,t)$ is the probability amplitude of finding the atom at 
the time $t$ at the position $x$ and the field in the $n$ Fock state.
Instead of this, we may apply the Wigner approach starting with proper full density matrix
$|\Psi(t)><\Psi(t)|$:
\begin{eqnarray}
\hat{\rho}=
\sum_{n',n''}w_{n'}w^*_{n''}\int\ud x'\int\ud x''
\Psi_{n'}(x',t)\Psi^*_{n''}(x'',t)|x'><x''|\otimes|n'><n''|.
\end{eqnarray}
Standard reduction gives pure atomic density matrix
\begin{eqnarray}
&&\hat{\rho}_a\equiv\int^{\infty}_{n=0}<n|\hat{\rho}|n>=\\
&&\sum|w_n|^2
\int\ud x'\int\ud x''\Psi_n(x',t)\Psi^*_n(x'',t)|x'><x''|\nonumber.
\end{eqnarray}
Then we have incoherent superposition 
\begin{equation}
W(x,p,t)=\sum^{\infty}_{n=0}|w_n|^2W_n(x,p,t)
\end{equation}
of the atomic Wigner functions (12)
corresponding to the atom motion in the potential $U_n(x)$ 
(which is not more than polynomial in $x$) generated by $n$-level Fock state. 
They are solutions of proper Wigner equations (13).
The next case describes the important decoherence process.
Let us have collective and environment subsystems with their own Hilbert spaces 
$\mathcal{H}=\mathcal{H}_c\otimes\mathcal{H}_e$
Relevant dynamics is described by three parts including interaction
$
H=H_c\otimes I_e+I_c\otimes H_e+H_{int}
$
For analysis, we can choose Lindblad master equation\cite{1}

\begin{eqnarray}
\dot{\rho}=\frac{1}{i\hbar}[H,\rho]-
\sum_n\gamma_n(L^+_nL_n\rho+
\rho L^+_nL_n-2L_n\rho L^+_n),
\end{eqnarray}
which preserves the positivity of density matrix and it is Markovian
but it is not general form of exact master equation.
Other choice is Wigner transform of master equation: 
\begin{eqnarray}
&&\dot{W}=\{H,W\}_{PB}+\\
&&\sum_{n\geq 1}\frac{\hbar^{2n}(-1)^n}{2^{2n}(2n+1)!}
\partial^{2n+1}_q U(q)\partial_p^{2n+1}W(q,p)+
2\gamma\partial_p pW+D\partial^2_pW\nonumber,
\end{eqnarray}
and it is more preferable for us.
In the next Section we consider the variational-wavelet approach Refs.~\refcite{2}--\refcite{7} 
for the solution of all
these Wigner-like equations (2), (8), (13), (20) for the case of an 
arbitrary polynomial $U(q, p)$, which corresponds to a finite number 
of terms in the series expansion in (13), (20) 
or to proper finite order of $\hbar$. Analogous approach can be 
applied to classical counterpart (5) also.
Roughly speaking,
wavelet analysis\cite{9} is some set of mathematical methods, which gives the possibility to
take into account high-localized states, control convergence of any type of expansions
and gives maximum sparse
forms for the general type of operators in such localized bases.
These bases are the natural generalization of standard coherent, 
squeezed, thermal squeezed states \cite{1},
which correspond to quadratic systems (pure linear dynamics) with Gaussian Wigner functions.
The representations of underlying symmetry group (affine group in the simplest case) 
on the proper functional space of states generate the exact multiscale expansions
which allow to control contributions to
the final result from each scale of resolution from the whole underlying 
infinite scale of spaces. 

\section{Variational Multiresolution Representation}

\subsection{Multiscale Decomposition for Space of States: 
Functional Realization and Metric Structure}

We obtain our multiscale/multiresolution representations for solutions of Wig\-ner-like equations
via a variational-wavelet approach. 
We represent the solutions as 
decomposition into localized eigenmodes (regarding action of affine group, i.e.
hidden symmetry of the underlying functional space of states) 
related to the hidden underlying set of scales: 
\begin{eqnarray}
W_n(t,q,p)=\displaystyle\bigoplus^\infty_{i=i_c}W^i_n(t,q,p),
\end{eqnarray}
where value $i_c$ corresponds to the coarsest level of resolution
$c$ or to the internal scale with the number $c$ in 
the full multiresolution decomposition (MRA)
of the underlying functional space ($L^2$, e.g.) corresponding 
to the problem under consideration\cite{9}:
\begin{equation}
V_c\subset V_{c+1}\subset V_{c+2}\subset\dots
\end{equation}
and $p=(p_1,p_2,\dots),\quad q=(q_1,q_2,\dots),\quad x_i=(p_1,q_1,\dots,p_i,q_i)$ 
are coordinates in phase space.
In the following we may consider as fixed as variable 
numbers of particles.
We introduce the Fock-like space structure (in addition to the standard one, 
if we consider second-quantized case) on the whole space of internal hidden scales
\begin{eqnarray}
H=\bigoplus_i\bigotimes_n H^n_i
\end{eqnarray}
for the set of n-partial Wigner functions (states):
\begin{equation}
W^i=\{W^i_0,W^i_1(x_1;t),\dots,
W^i_N(x_1,\dots,x_N;t),\dots\},
\end{equation}
where
$W_p(x_1,\dots, x_p;t)\in H^p$,
$H^0=C,\quad H^p=L^2(R^{6p})$ (or any different proper functional spa\-ce), 
with the natural Fock space like norm: 
\begin{eqnarray}
(W,W)=W^2_0+
\sum_{i}\int W^2_i(x_1,\dots,x_i;t)\prod^i_{\ell=1}\mu_\ell.
\end{eqnarray}
First of all, we consider $W=W(t)$ as a function of time only,
$W\in L^2(R)$, via
multiresolution decomposition which naturally and efficiently introduces 
the infinite sequence of the underlying hidden scales \cite{9}.
We have the contribution to
the final result from each scale of resolution from the whole
infinite scale of spaces (22).
The closed subspace
$V_j (j\in {\bf Z})$ corresponds to  the level $j$ of resolution, 
or to the scale j
and satisfies
the following properties:
let $D_j$ be the orthonormal complement of $V_j$ with respect to $V_{j+1}$: 
$
V_{j+1}=V_j\bigoplus D_j.
$
Then we have the following decomposition:
\begin{eqnarray}
\{W(t)\}=\bigoplus_{-\infty<j<\infty} D_j 
=\overline{V_c\displaystyle\bigoplus^\infty_{j=0} D_j},
\end{eqnarray}
in case when $V_c$ is the coarsest scale of resolution.
The subgroup of translations generates a basis for the fixed scale number:
$
{\rm span}_{k\in Z}\{2^{j/2}\Psi(2^jt-k)\}=D_j.
$
The whole basis is generated by action of the full affine group:
\begin{eqnarray}
{\rm span}_{k\in Z, j\in Z}\{2^{j/2}\Psi(2^jt-k)\}=
{\rm span}_{k,j\in Z}\{\Psi_{j,k}\}
=\{W(t)\}.
\end{eqnarray}

\subsection{Tensor Product Structure}

Let sequence 
$\{V_j^t\}, V_j^t\subset L^2(R)$
correspond to multiresolution analysis on time axis and 
$
\{V_j^{x_i}\},\qquad V_j^{x_i}\subset L^2(R)
$
correspond to multiresolution analysis for coordinate $x_i$,
then
$
V_j^{n+1}=V^{x_1}_j\otimes\dots\otimes V^{x_n}_j\otimes  V^t_j
$
corresponds to multiresolution analysis for n-particle distribution function 
$W_n(x_1,\dots,x_n;t)$.
E.g., for $n=2$:
\begin{eqnarray}
V^2_0=\{f:f(x_1,x_2)=
\sum_{k_1,k_2}a_{k_1,k_2}\phi^2(x_1-k_1,x_2-k_2),\quad
a_{k_1,k_2}\in\ell^2(Z^2)\},
\end{eqnarray}
where 
$
\phi^2(x_1,x_2)=\phi^1(x_1)\phi^2(x_2)=\phi^1\otimes\phi^2(x_1,x_2),
$
and $\phi^i(x_i)\equiv\phi(x_i)$ form a multiresolution basis corresponding to
$\{V_j^{x_i}\}$.
If 
$
\{\phi^1(x_1-\ell)\},\ \ell\in Z
$
 form an orthonormal set, then 
$
\phi^2(x_1-k_1, x_2-k_2)
$
form an orthonormal basis for $V^2_0$.
Action of affine group provides us by multiresolution representation of
$L^2(R^2)$. After introducing detail spaces $D^2_j$, we have, e.g. 
$
V^2_1=V^2_0\oplus D^2_0.
$
Then
3-component basis for $D^2_0$ is generated by translations of three functions 
\begin{equation}
\Psi^2_1=\phi^1(x_1)\otimes\Psi^2(x_2),\
\Psi^2_2=\Psi^1(x_1)\otimes\phi^2(x_2),\
\Psi^2_3=\Psi^1(x_1)\otimes\Psi^2(x_2).
\end{equation}
Also, we may use the rectangle lattice of scales and one-dimensional wavelet
decomposition :
\begin{equation}
f(x_1,x_2)=\sum_{i,\ell;j,k}<f,\Psi_{i,\ell}\otimes\Psi_{j,k}>
\Psi_{j,\ell}\otimes\Psi_{j,k}(x_1,x_2),
\end{equation}
where bases functions
$ 
\Psi_{i,\ell}\otimes\Psi_{j,k}
$
depend on two scales $2^{-i}$ and $2^{-j}$.

After construction the multidimensional bases  
we obtain our multiscale\-/mul\-ti\-re\-so\-lu\-ti\-on 
representations for observables (symbols), states, partitions
via the variational approaches in Refs.~\refcite{2}--\refcite{7} as for c-BBGKY as for its quantum counterpart
and related reductions but before we need to construct reasonable multiscale 
decomposition for all operators included in the set-up.

\subsection{FWT Decomposition for Observables}

One of the key point of wavelet analysis approach, 
the so called Fast Wavelet Transform (FWT) \cite{9}, 
demonstrates that for the large classes of
operators the wavelet-like functions are best approximation for true eigenvectors and the corresponding 
matrices are almost diagonal. So, powerful FWT provides the maximum sparse form for different classes 
of operators \cite{9}.
Let us denote our (integral/differential) operator from equations under 
consideration  
as  $T$ ($L^2(R^n)\rightarrow L^2(R^n)$) and its kernel as $K$.
We have the following representation:
\begin{equation}
<Tf,g>=\int\int K(x,y)f(y)g(x)\ud x\ud y.
\end{equation}
In case when $f$ and $g$ are wavelets $\varphi_{j,k}=2^{j/2}\varphi(2^jx-k)$, 
(21) provides the standard representation for operator $T$.
Let us consider multiresolution representation
$
\dots\subset V_2\subset V_1\subset V_0\subset V_{-1}
\subset V_{-2}\dots
$. 
The basis in each $V_j$ is 
$\varphi_{j,k}(x)$,
where indices $\ k, j$ represent translations and scaling 
respectively. 
Let $P_j: L^2(R^n)\rightarrow V_j$ $(j\in Z)$ be projection
operators on the subspace $V_j$ corresponding to level $j$ of resolution:
$
(P_jf)(x)=\sum_k<f,\varphi_{j,k}>\varphi_{j,k}(x).
$ 
Let
$Q_j=P_{j-1}-P_j$ be the projection operator on the subspace $D_j$ ($V_{j-1}=V_j\oplus D_j$), 
then
we have the following 
representation of operator T which takes into account contributions from
each level of resolution from different scales starting with the
coarsest and ending to the finest scales \cite{9}:
\begin{equation}
T=\sum_{j\in Z}(Q_jTQ_j+Q_jTP_j+P_jTQ_j).
\end{equation}
We need to remember that this is a result of presence of affine group inside this
construction.
The non-standard form of operator representation is a representation of
operator T as  a chain of triples
$T=\{A_j,B_j,\Gamma_j\}_{j\in Z}$, acting on the subspaces $V_j$ and
$D_j$:
$
 A_j: D_j\rightarrow D_j, B_j: V_j\rightarrow D_j,
\Gamma_j: D_j\rightarrow V_j,
$
where operators $\{A_j,B_j,\Gamma_j\}_{j\in Z}$ are defined
as
$A_j=Q_jTQ_j, \quad B_j=Q_jTP_j, \quad\Gamma_j=P_jTQ_j.$
The operator $T$ admits a recursive definition via
\begin{eqnarray}
T_j=
\left(\begin{array}{cc}
A_{j+1} & B_{j+1}\\
\Gamma_{j+1} & T_{j+1}
\end{array}\right),
\end{eqnarray}
where $T_j=P_jTP_j$ and $T_j$ acts on $ V_j: V_j\rightarrow V_j$.
So, it is possible to provide the following ``sparse'' action of operator $T_j$
on elements $f$ of functional realization of our space of states $H$:
\begin{equation}
(T_j f)(x)=\sum_{k\in Z}\left(2^{-j}\sum_{\ell}r_\ell f_{j,k-\ell}\right)
\varphi_{j,k}(x),
\end{equation}
in the wavelet basis $\varphi_{j,k}(x)=2^{-j/2}\varphi(2^{-j}x-k)$, where
\begin{equation}
f_{j,k-1}=2^{-j/2}\int f(x)\varphi(2^{-j}x-k+\ell)\ud x
\end{equation}
are wavelet coefficients and $r_\ell$  
are the roots of some additional linear system of equations related to
the ``type of localization'' \cite{9}.
So, we have the simple linear para\-met\-rization of
matrix representation of  our operators in localized wavelet bases
and of the action of
this operator on arbitrary vector/state in proper functional space.

\subsection{Variational Approach}

Now, after preliminary work with (functional) spaces, states and operators, 
we may apply our variational approach from [2]-[7].
Let $L$ be an arbitrary (non)li\-ne\-ar dif\-fe\-ren\-ti\-al\-/\-in\-teg\-ral operator 
 with matrix dimension $d$
(finite or infinite), 
which acts on some set of functions
from $L^2(\Omega^{\otimes^n})$:  
$\quad\Psi\equiv\Psi(t,x_1,x_2,\dots)=\Big(\Psi^1(t,x_1,x_2,\dots), \dots$,
$\Psi^d(t,x_1,x_2,\dots)\Big)$,
 $\quad x_i\in\Omega\subset{\bf R}^6$, $n$ is the number of particles:
\begin{eqnarray}
L\Psi&\equiv& L(Q,t,x_i)\Psi(t,x_i)=0,\\
Q&\equiv& Q_{d_0,d_1,d_2,\dots}(t,x_1,x_2,\dots,
\partial /\partial t,\partial /\partial x_1,
\partial /\partial x_2,\dots,
\int \mu_k)\nonumber\\
&=&
\sum_{i_0,i_1,i_2,\dots=1}^{d_0,d_1,d_2,\dots}
q_{i_0i_1i_2\dots}(t,x_1,x_2,\dots)
\Big(\frac{\partial}{\partial t}\Big)^{i_0}\Big(\frac{\partial}{\partial x_1}\Big)^{i_1}
\Big(\frac{\partial}{\partial x_2}\Big)^{i_2}\dots\int\mu_k.\nonumber 
\end{eqnarray}
Let us consider now the $N$ mode approximation for the solution as 
the following ansatz:
\begin{eqnarray}
\Psi^N(t,x_1,x_2,\dots)=
\sum^N_{i_0,i_1,i_2,\dots=1}a_{i_0i_1i_2\dots}
 A_{i_0}\otimes 
B_{i_1}\otimes C_{i_2}\dots(t,x_1,x_2,\dots)
\end{eqnarray}
We will determine the expansion coefficients from the following conditions
(related to proper choosing of variational approach):
\begin{eqnarray}
&&\ell^N_{k_0,k_1,k_2,\dots}\equiv 
\int(L\Psi^N)A_{k_0}(t)B_{k_1}(x_1)C_{k_2}(x_2)\ud t\ud x_1\ud x_2\dots=0.
\end{eqnarray}
Thus, we have exactly $dN^n$ algebraical equations for  $dN^n$ unknowns 
$a_{i_0,i_1,\dots}$.
This variational ap\-proach reduces the initial problem 
to the problem of solution 
of functional equations at the first stage and 
some algebraical problems at the second one.
It allows to unify the multiresolution expansion with variational 
construction in Refs.~\refcite{2}--\refcite{7}. 

As a result, the solution is parametrized by the solutions of two sets of 
reduced algebraical
problems, one is linear or nonlinear
(depending on the structure of the generic operator $L$) and the rest are linear
problems related to the computation of the coefficients of reduced 
algebraic equations. It is also related to the choice of exact measure of localization
(including class of smoothness) which are proper for our set-up.
These coefficients can be found  by some functional/algebraic methods
by using the
compactly supported wavelet basis functions or any other wavelet families \cite{9}.
As a result the solution of the equations/hierarchies from Section 4, as in c-
as in q-region, has the 
following mul\-ti\-sca\-le or mul\-ti\-re\-so\-lu\-ti\-on decomposition via 
nonlinear high\--lo\-ca\-li\-zed eigenmodes 
{\setlength\arraycolsep{0pt}
\begin{eqnarray}
&&W(t,x_1,x_2,\dots)=
\sum_{(i,j)\in Z^2}a_{ij}U^i\otimes V^j(t,x_1,\dots),\nonumber\\
&&V^j(t)=
V_N^{j,slow}(t)+\sum_{l\geq N}V^j_l(\omega_lt), \ \omega_l\sim 2^l, 
\end{eqnarray}
}
$$U^i(x_s)=
U_M^{i,slow}(x_s)+\sum_{m\geq M}U^i_m(k^{s}_mx_s), \ k^{s}_m\sim 2^m,
$$
which corresponds to the full multiresolution expansion in all underlying time/space 
scales.
The formulae (39) give the expansion into a slow part
and fast oscillating parts for arbitrary $N, M$.  So, we may move
from the coarse scales of resolution to the 
finest ones for obtaining more detailed information about the dynamical process.
In this way one obtains contributions to the full solution
from each scale of resolution or each time/space scale or from each nonlinear eigenmode.
It should be noted that such representations 
give the best possible localization
properties in the corresponding (phase)space/time coordinates. 
Formulae (39) do not use perturbation
techniques or linearization procedures.
Numerical calculations are based on compactly supported
wavelets and wavelet packets and on evaluation of the 
accuracy on 
the level $N$ of the corresponding cut-off of the full system 
regarding norm (25):
$
\|W^{N+1}-W^{N}\|\leq\varepsilon.
$

\section{Modeling of Patterns}

To summarize, the key points are:

1. The ansatz-oriented choice of the (multi\-di\-men\-si\-o\-nal) ba\-ses
related to some po\-ly\-no\-mi\-al tensor algebra. 

2. The choice of proper variational principle. A few 
pro\-je\-c\-ti\-on/ \-Ga\-ler\-kin\--li\-ke 
principles for constructing (weak) solutions can be considered.
The advantages of formulations related to biorthogonal
(wavelet) decomposition should be noted. 

3. The choice of  bases functions in the scale spaces $D_j$ from wavelet zoo. They 
correspond to high-localized (nonlinear) excitations, 
nontrivial local (stable) distributions/fluctuations or ``continuous qudits''.
Besides fast convergence properties it should be noted 
minimal complexity of all underlying calculations, especially in case of choice of wavelet
packets which minimize Shannon entropy. 

4.  Operator  representations providing maximum sparse representations 
for arbitrary (pseudo) differential/ integral operators 
$\ud f/\ud x$, $\ud^n f/\ud x^n$, $\int T(x,y)f(y)\ud y)$, etc.

5. (Multi)linearization. Besides the variation approach we can consider also a different method
to deal with (polynomial) nonlinearities: para-products-like decompositions.

To classify the qualitative behaviour we apply
standard methods from general control theory or really use the 
control.
We will start from a priori unknown coefficients, the exact values of which 
will subsequently be recovered.
Roughly speaking, we will fix only class of nonlinearity 
(polynomial in our case)
which covers a broad variety of examples of possible truncation of the systems.
As a simple model we choose band-triangular 
non-sparse matrices $(a_{ij})$.
These matrices provide tensor structure of bases in (extended) phase space
and are generated by the roots of the reduced variational (Galerkin-like) 
systems.
As a second step we need to restore the coefficients from these 
matrices
by which we may classify the types of behaviour. 
We start with the localized mode, which is a base mode/eigenfunction, 
which was constructed as a tensor product of the two base functions. 
Fig.~8, 11 below demonstrate 
the result of summation of series (39) up to value of the 
dilation/scale parameter equal to four and six, respectively.
It's done in the bases of symmlets [9] with the corresponding matrix 
elements equal to one.  The size of matrix of ``Fourier-wavelet coefficients''
is 512x512. So, different possible  distributions of  the root values 
of the generic 
algebraical systems (38) provide qualitatively different types of behaviour.
Generic algebraic system (38), Generalized Dispersion Relation (GDR), 
provide the possibility for algebraic control. 
The above choice
provides us by a distribution with chaotic-like equidistribution.
But, if we consider a band-like structure of matrix $(a_{ij})$ 
with the band along the main diagonal with 
finite size ($\ll 512$) and values, e.g. five, while the other 
values are equal to one, we obtain
localization in a fixed finite area of the full phase space, 
i.e. almost all energy of the system 
is concentrated in this small volume. This corresponds to waveleton states \cite{7}
and 
is shown in Fig.~9, constructed by means of Daubechies-based wavelet packets. 
Depending on the type of 
solution,  such localization may be conserved during the whole time evolution
(asymptotically-stable) or up to the needed value from the whole time scale (e.g. enough 
for plasma fusion/confinement in the case of fusion modeling by means of c-BBGKY hierarchy
for dynamics of partitions).

\section{Conclusions}

By using wavelet bases with their best phase space      
localization  properties, we can describe the localized (coherent) structures in      
quantum systems with complicated behaviour (Figs.~8, 11).
The numerical simulation demonstrates the formation of different (stable) pattern or orbits 
generated by internal hidden symmetry from
high-localized structures.
Our (nonlinear) eigenmodes are more realistic for the modeling of 
nonlinear classical/quantum dynamical process  than the corresponding linear gaussian-like
coherent states. Here we mention only the best convergence properties of the expansions 
based on wavelet packets, which  realize the minimal Shannon entropy property
and the exponential control of convergence of expansions like (39) 
based on the norm (25).
 Fig.~9 corresponds to (possible) result of superselection
(einselection) [1] after decoherence process started from entangled state (Fig.~12);
Fig.~10 and Fig.~13 demonstrate the steps of multiscale resolution 
(or degrees of interference) during modeling (quantum interaction/evolution)  
of entangled states leading to the growth of degree of entanglement.
It should be noted that
we can control the type of behaviour on the level of the reduced algebraical variational 
system, GDR (38). 

Let us finish with some phenomenological description which can be 
considered as 
an attempt of qualitative description of the quantum dynamics 
as a whole and in comparison with its classical counterpart.
It is possible to take for reminiscence the famous Dirac's phrase that ``an electron can interact only 
itself via the process of quantum interference''.
Let $G$ be the hidden/internal symmetry group on the spaces of quantum states which generates via MRA
(22), (26) the multiscale/multiresolution representation for all 
dynamical quantities, unified in object $O(t)$, such as states, observables, partitions: 
$O^i(t)=\{\psi^i(t), Op^i(t), W_n^i(t)\}$, where $i$ is the proper scale index.
Then, the following commutative diagram represents the details of quantum life from 
the point of view of representations of $G$ on the chosen functional realization which leads to
decomposition of the whole quantum evolution into the proper orbits or scales corresponding 
to the proper level of resolution. Morphisms $W(t)$ describe Wigner-Weyl evolution in the 
algebra of symbols, while the processes of interactions with open World, such as
the measurement or decoherence, correspond to morphisms (or even functors) $m(t)$
which transform the infinite set of scales
characterizing the quantum object into finite ones, sometimes consisting of one element
(demolition/destructive measurement).
\begin{eqnarray}
&&\qquad\qquad\qquad W(t) \nonumber\\
&&\lbrace O^i(t_1) \rbrace \qquad  \longrightarrow   \qquad \lbrace O^j(t_2) \rbrace \nonumber\\
&&\qquad\qquad\nonumber\\
&&\downarrow m(t_1) \qquad \qquad \qquad  \downarrow m(t_2) \nonumber\\
&&\qquad\qquad\qquad \widetilde{W(t)} \nonumber\\
&&\lbrace O^{i_c}(t_1) \rbrace \qquad  \longrightarrow  \qquad \lbrace O^{j_c}(t_2) \rbrace,\nonumber
\end{eqnarray}
where reduced morphisms $\widetilde{W(t)}$ correspond to (semi)classical or 
quasiclassical evolution. 
So, qualitatively,

{\bf Quantum Objects} can be represented by an infinite or sufficiently large set of coexisting and
interacting subsets like (22), (26) 
while 

{\bf Classical Objects} can be described by one or few only levels of resolution
with (almost) suppressed interscale self-interaction.
It is possible to consider Wigner functions as some measure of the quantum character of the system:
as soon as it becomes positive, we arrive to classical regime and so there is 
no need to consider the full hierarchy decomposition in the representation 
(21).

So, Dirac's self-interference is nothing but the multiscale mixture/intermittency.
Certainly, the degree of this self-interaction leads to different qualitative types 
of behaviour, such as localized quasiclassical states, separable, entangled, chaotic etc.
At the same time the instantaneous quantum interaction or transmission of 
(quantum) information from Alice 
to Bob takes place not in the physical kinematical space-time but in Hilbert spaces of states
in their proper functional realization where there is a different kinematic life.
To describe a set of Quantum Objects we need to realize our Space of States (Hilbert space)
not as one functional space but as the so-called and well known in 
mathematics scale of spaces, e.g. $B^s_{p,q}$, $F^s_{p,q}$ \cite{9}.
The proper multiscale decomposition for the scale of space provides us by the 
method of description of the set of quantum objects in case if the ``size'' of 
one Hilbert space of states is not enough to describe the complicated internal World.
We will consider it elsewhere, while here we considered the one-scale case
(to avoid possible misunderstanding we need to mention that one-scale case 
is also described by an infinite scale of spaces (26), but it is internal 
decomposition of the unique, attached to the problem, Hilbert space).

\section*{Acknowledgements}

We are very grateful to Prof. M. Planat for kind help, nice hospitality and patience.

\newpage

\begin{twocolumn}

\begin{figure}
\begin{center}
\begin{tabular}{c}
\includegraphics[width=55mm]{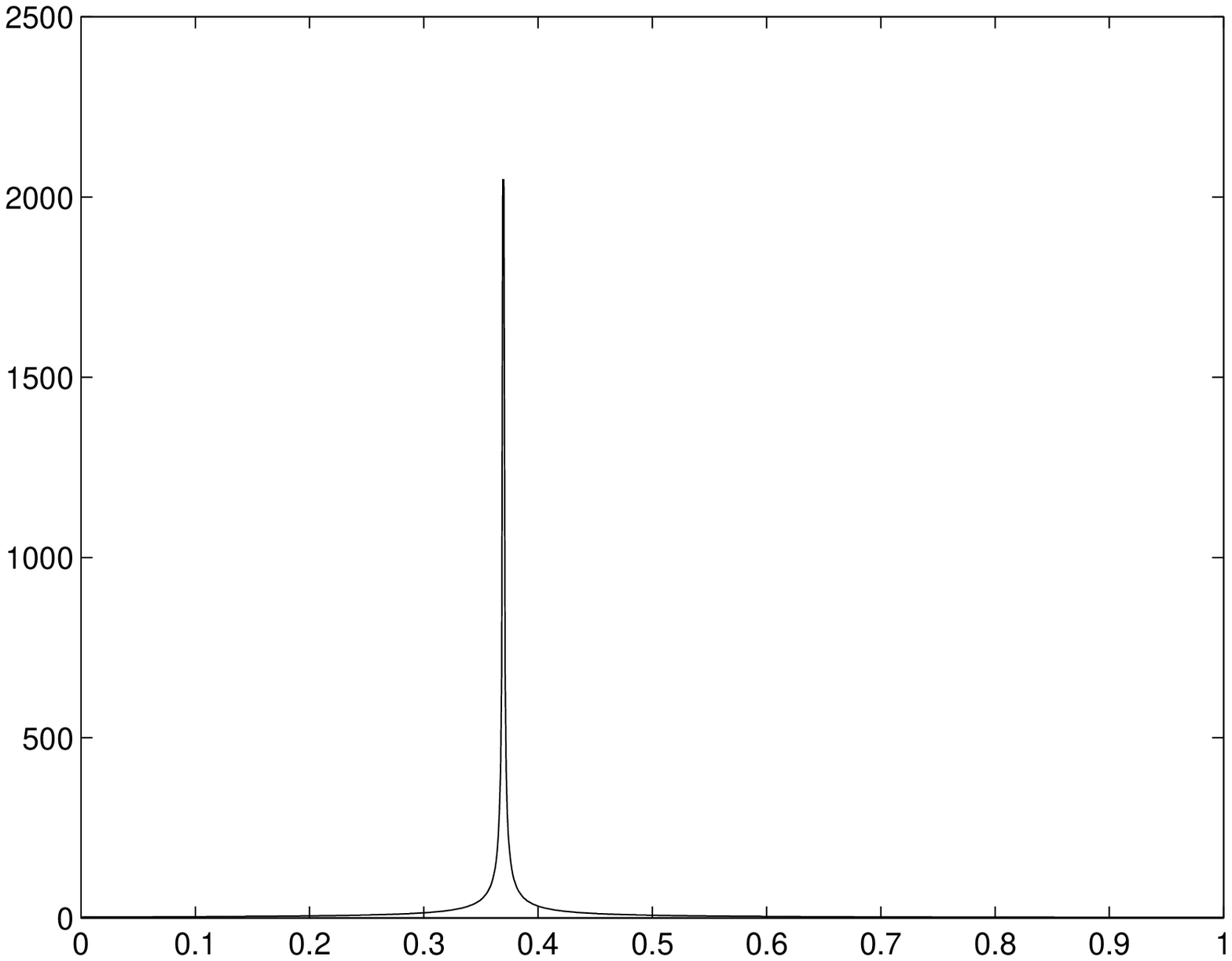}
\end{tabular}
\end{center}
\caption{Kick.}
\end{figure}

\begin{figure}
\begin{center}
\begin{tabular}{c}
\includegraphics[width=55mm]{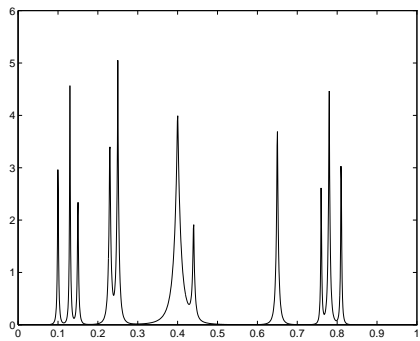}
\end{tabular}
\end{center}
\caption{Multi-Kicks.}
\end{figure}

\begin{figure}
\begin{center}
\begin{tabular}{c}
\includegraphics[width=55mm]{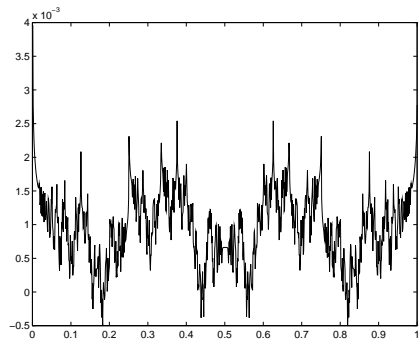}
\end{tabular}
\end{center}
\caption{RW-fractal.}
\end{figure}

\begin{figure}
\begin{center}
\begin{tabular}{c}
\includegraphics[width=55mm]{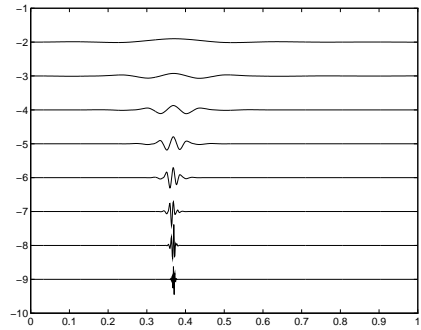}
\end{tabular}
\end{center}
\caption{MRA for Kick.}
\end{figure}

\begin{figure}
\begin{center}
\begin{tabular}{c}
\includegraphics[width=55mm]{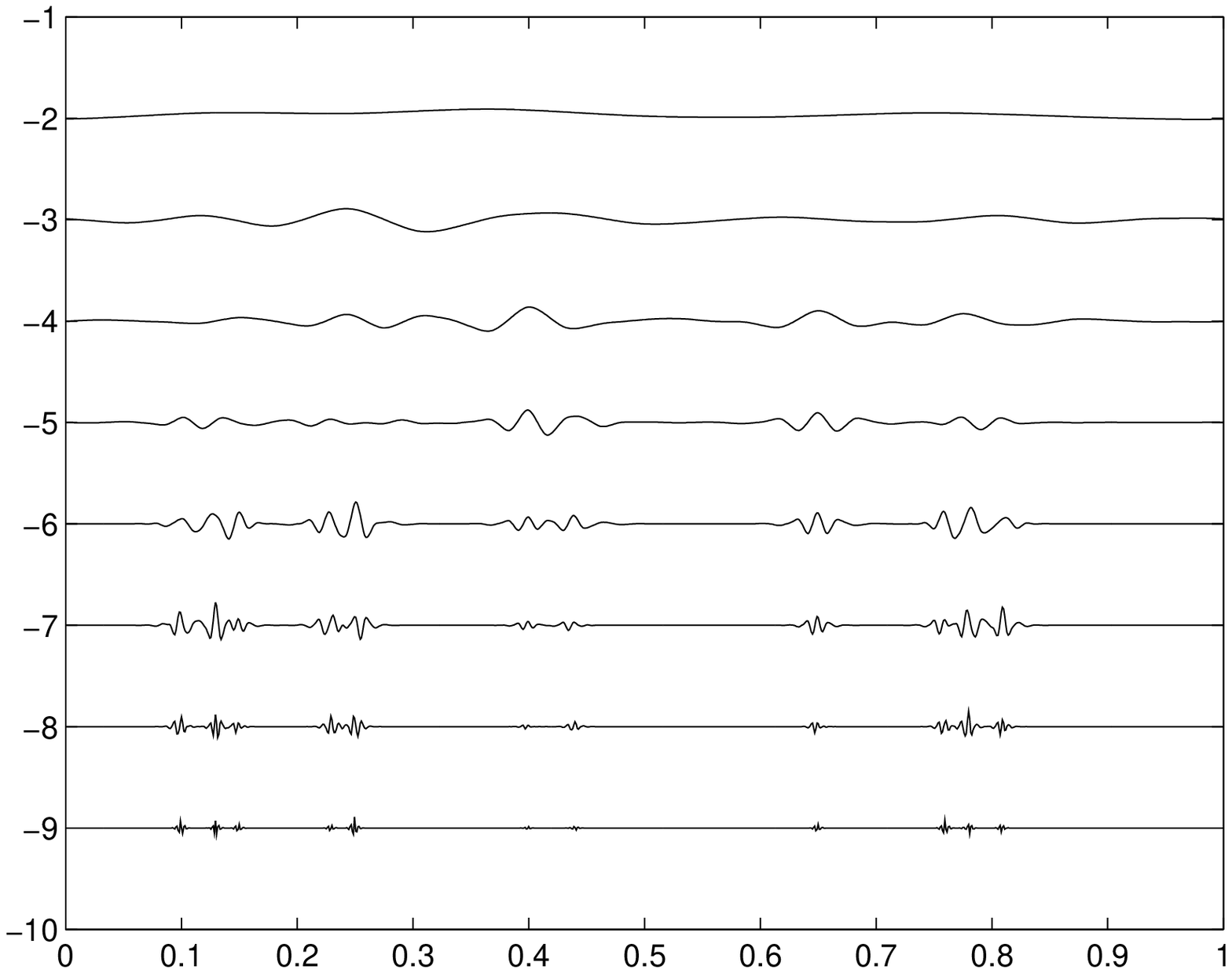}
\end{tabular}
\end{center}
\caption{MRA for Multi-Kicks.}
\end{figure}

\begin{figure}
\begin{center}
\begin{tabular}{c}
\includegraphics[width=55mm]{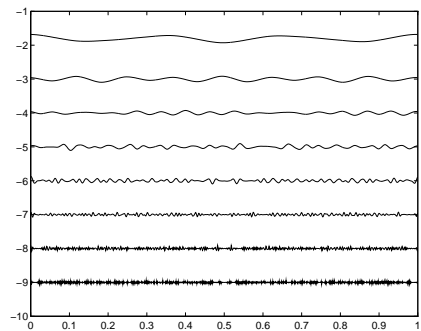}
\end{tabular}
\end{center}
\caption{MRA for RW-fractal.}
\end{figure}

\end{twocolumn}

\twocolumn
\newpage

\begin{figure}
\begin{center}
\begin{tabular}{c}
\includegraphics*[width=50mm]{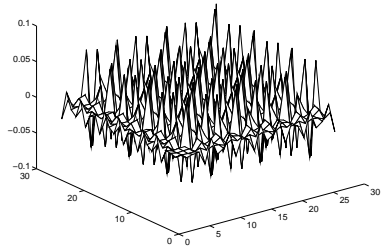}
\end{tabular}
\end{center}
\caption{Level 4 MRA.}
\end{figure}

\vspace*{10mm}

\begin{figure}
\begin{center}
\begin{tabular}{c}
\includegraphics*[width=50mm]{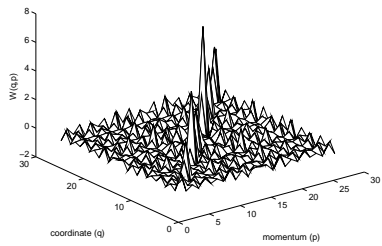}
\end{tabular}
\end{center}
\caption{Localized pattern, (waveleton) Wigner function.}
\end{figure}

\vspace*{5mm}

\begin{figure}
\begin{center}
\begin{tabular}{c}
\includegraphics*[width=50mm]{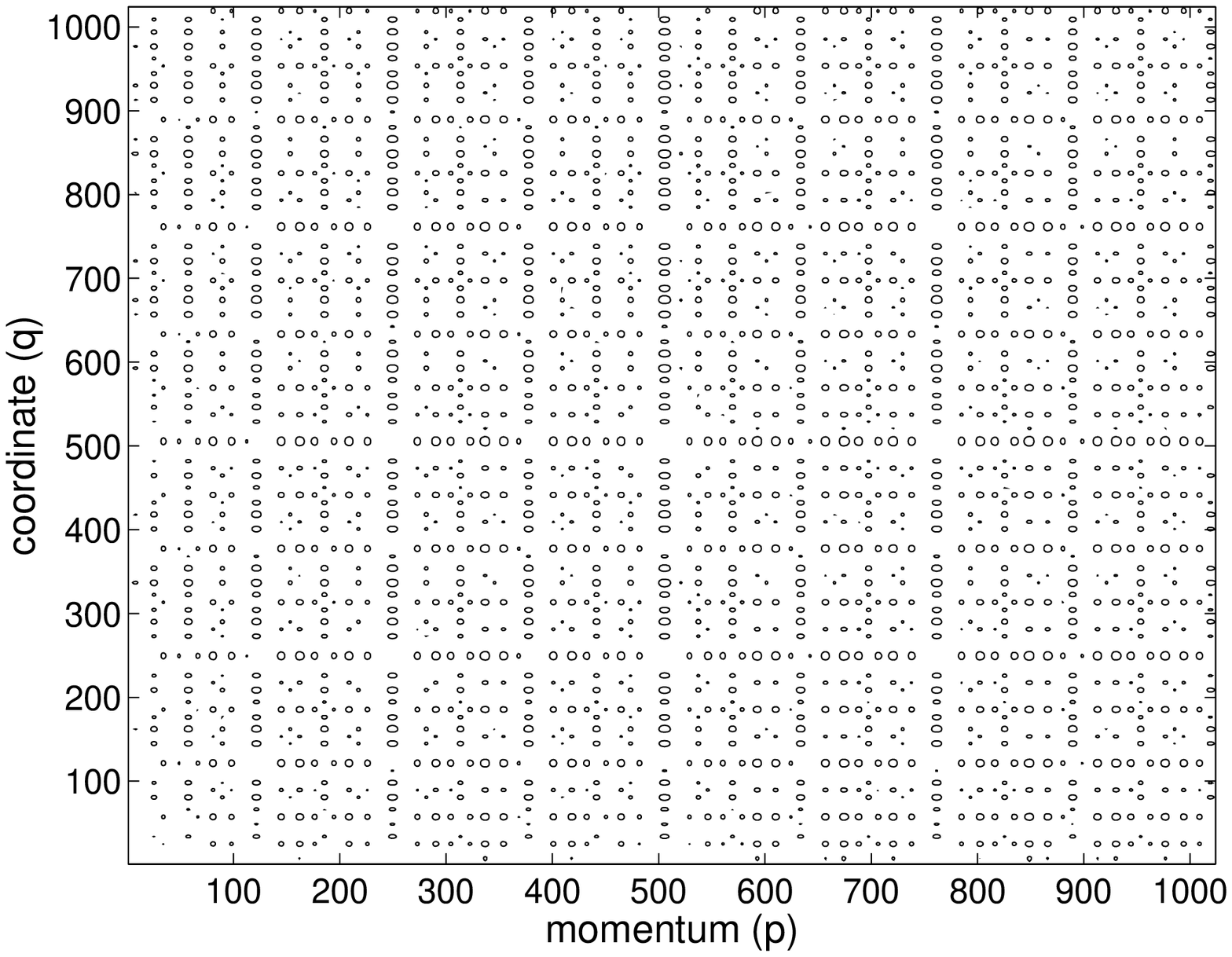}
\end{tabular}
\end{center}
\caption{Interference picture on the level 4 approximation for Wigner function.}
\end{figure}

\begin{figure}
\begin{center}
\begin{tabular}{c}
\includegraphics*[width=50mm]{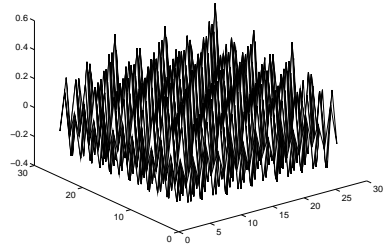}
\end{tabular}
\end{center}
\caption{Level 6 MRA.}
\end{figure}

\begin{figure}
\begin{center}
\begin{tabular}{c}
\includegraphics*[width=50mm]{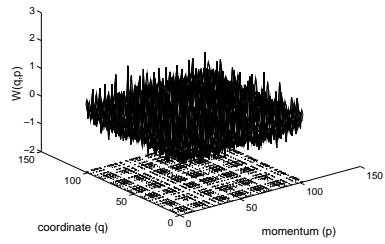}
\end{tabular}
\end{center}
\caption{Entangled-like Wigner function.}
\end{figure}

\begin{figure}
\begin{center}
\begin{tabular}{c}
\includegraphics*[width=50mm]{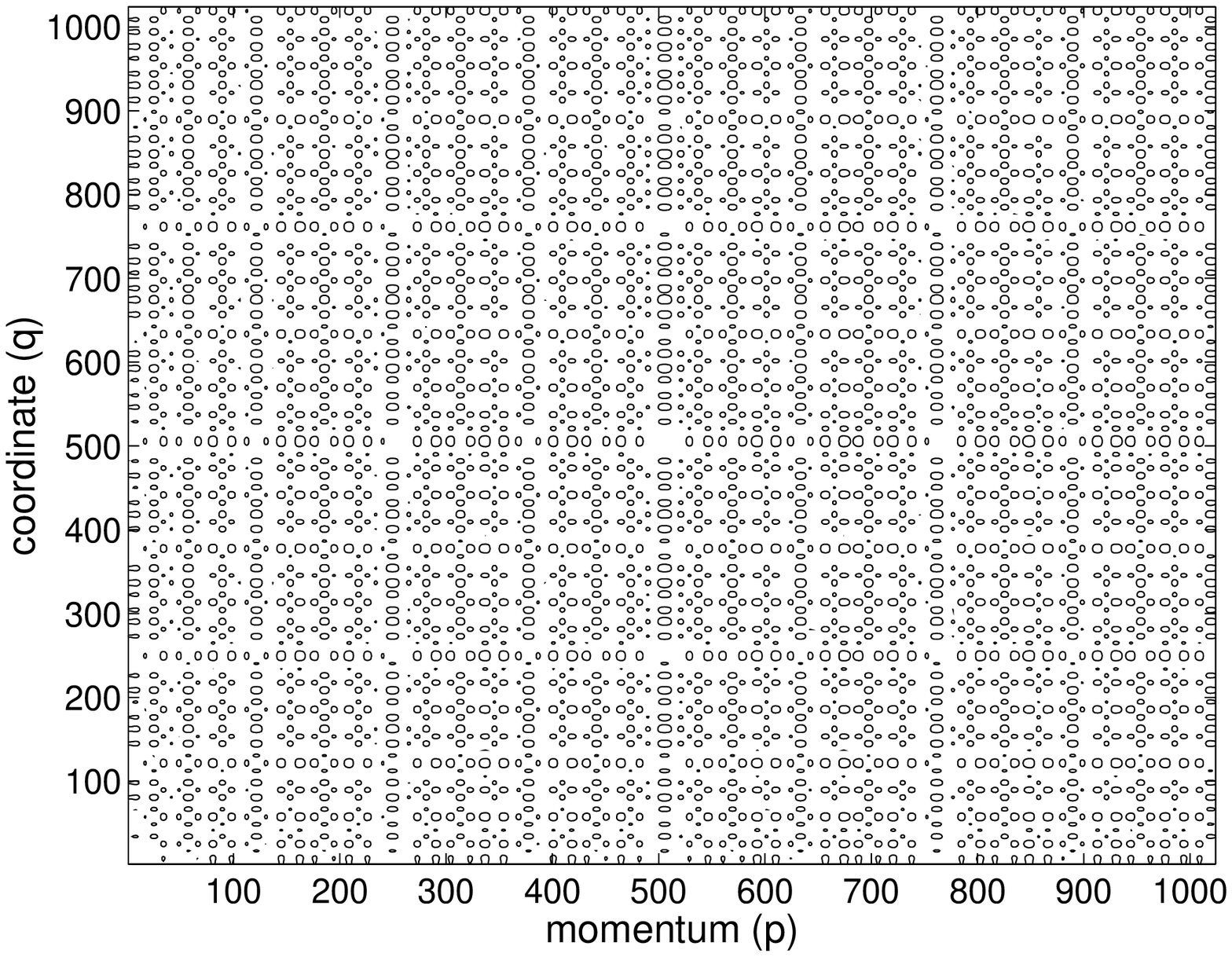}
\end{tabular}
\end{center}
\caption{Interference picture on the level 6 approximation for Wigner function.}
\end{figure}

\onecolumn

\end{document}